%% file: main.tex
\documentclass[sigconf]{acmart} % ,authordraft,anonymous=true
\usepackage[linesnumbered,vlined,ruled]{algorithm2e}
\usepackage{amsmath}
\usepackage{cleveref}
\usepackage{caption}
\usepackage{subcaption}
\usepackage{tabularx}
\usepackage{numprint}
\usepackage{graphicx}
\usepackage{capt-of}% or \usepackage{caption}
\npthousandsep{,}\npthousandthpartsep{}\npdecimalsign{.}

\AtBeginDocument{%
  \providecommand\BibTeX{{%
    \normalfont B\kern-0.5em{\scshape i\kern-0.25em b}\kern-0.8em\TeX}}}

\copyrightyear{2021}
\acmYear{2021}
\setcopyright{rightsretained}
\acmConference[CIKM '21]{Proceedings of the 30th ACM International Conference on Information and Knowledge Management}{November 1--5, 2021}{Virtual Event, QLD, Australia}
\acmBooktitle{Proceedings of the 30th ACM International Conference on Information and Knowledge Management (CIKM '21), November 1--5, 2021, Virtual Event, QLD, Australia}
\acmDOI{10.1145/3459637.3482318}
\acmISBN{978-1-4503-8446-9/21/11}
\newcommand\amppere{\textbf{AMPPERE} }

\begin{document}
\fancyhead{}

\title{AMPPERE: A Universal Abstract Machine for Privacy-Preserving Entity Resolution Evaluation}

%\numberofauthors{4}

\author{Yixiang Yao}
\email{yixiangy@usc.edu}
\orcid{0000-0002-2471-5591}
\affiliation{%
  \institution{USC \& Information Sciences Institute}
  \streetaddress{}
  \city{Marina del Rey}
  \state{California}
  \country{USA}
  \postcode{90292}
}

\author{Tanmay Ghai}
\email{tghai@usc.edu}
\orcid{0000-0001-7200-3364}
\affiliation{%
  \institution{USC \& Information Sciences Institute}
  \streetaddress{}
  \city{Marina del Ray}
  \state{California}
  \country{USA}
  \postcode{90089}
}

\author{Srivatsan Ravi}
\email{srivatsr@usc.edu}
\orcid{0000-0002-2965-3940}
\affiliation{%
  \institution{USC \& Information Sciences Institute}
  \streetaddress{}
  \city{Marina del Rey}
  \state{California}
  \country{USA}
  \postcode{90292}
}

\author{Pedro Szekely}
\email{szekely@usc.edu}
\orcid{0000-0002-4621-2266}
\affiliation{%
  \institution{USC \& Information Sciences Institute}
  \streetaddress{}
  \city{Marina del Rey}
  \state{California}
  \country{USA}
  \postcode{90292}
}

% \renewcommand{\shortauthors}{Anonymous, et al.}

\begin{abstract}
Entity resolution is the task of identifying records in different datasets that refer to the same entity in the real world.
In sensitive domains (e.g. financial accounts, hospital health records), entity resolution must meet privacy requirements to avoid revealing sensitive information such as personal identifiable information to untrusted parties.
Existing solutions are either too algorithmically-specific or come with an implicit trade-off between accuracy of the computation, privacy, and run-time efficiency. 
We propose \textbf{AMMPERE}, an abstract computation model for performing \emph{universal} privacy-preserving entity resolution. \amppere offers abstractions that encapsulate multiple algorithmic and platform-agnostic approaches using variants of Jaccard similarity to perform private data matching and entity resolution. Specifically, we show that two parties can perform entity resolution over their data, without leaking sensitive information. We rigorously compare and analyze the feasibility, performance overhead and privacy-preserving properties of these approaches on the Sharemind multi-party computation (MPC) platform as well as on PALISADE, a lattice-based homomorphic encryption library. The \amppere system demonstrates the efficacy of privacy-preserving entity resolution for real-world data while providing a precise characterization of the induced cost of preventing information leakage.
\end{abstract}

% %%
% %% The code below is generated by the tool at http://dl.acm.org/ccs.cfm.
% %% Please copy and paste the code instead of the example below.
% %%



\maketitle

\setlength{\textfloatsep}{1pt}% Remove margin of algorithms, tables, graphs, etc

\input{intro.tex}

\input{preliminary.tex}

\input{approach.tex}

\input{experiment.tex}

\input{relatedwork.tex}

\input{conclusion.tex}

\begin{acks}
This project was supported by the DataSafes program with funding from Actuate and the Alfred P. Sloan Foundation.
\end{acks}

\balance
\bibliographystyle{ACM-Reference-Format}
\bibliography{ref}

\end{document}

%% file: intro.tex
\section{Introduction}
% 0.5-0.75 page.
% \begin{itemize}
%     \item Entity resolution
%     \item Sensitive data scenarios: e.g., patient info from hospital
%     \item Privacy preserving entity resolution
%     \item Enhancing privacy with FHE and Secure MPC.
% \end{itemize}
In the real world, performing entity resolution over data from varied and distributed data sources is a challenging problem. For example, consider two different patient databases belonging to different hospitals, the first one containing records for Peter and Bruce, and the second for Tony, Pet, and Brvce (Pet is an abbreviation for Peter \& Brvce is a misspelling of Bruce). 

The \emph{entity resolution} (ER) task here is to identify records from the two databases that belong to the same patients. An ER algorithm will compare the information present in the records (e.g. names, home or work address, phone numbers), and determine if the records refer to the same person. The principal challenge with this is that the two hospitals may record patient information in different formats (i.e. the first may use separate fields for first and last name, whereas the second may store the full name as one field). The \emph{privacy-preserving entity resolution} (PPER) task is an extension of the ER task, with the additional requirement to not reveal sensitive information to any party present in the computation or to an adversary. Specifically, each party learns which of its records are present on other parties, but nothing about any other records; an adversary should learn nothing about \emph{any} records of \emph{any} party. Illustrating this with our original example, the first hospital may learn that the second hospital has records for Peter and Bruce, but should not learn about a record for Tony, or any other information about the patients themselves unless explicitly disclosed by the second hospital. 
% Furthermore, there should be nothing about the distribution of names, locations, or any other relevant information that the first hospital should learn about the second.
Likewise, the second hospital should \emph{only} learn about Peter \& Bruce's records in the first hospital's database. 

Privacy preservation adds significant challenges and computational overhead in solving ER as the data must be obfuscated to preserve privacy, making similarity comparisons difficult and costly. This is especially the case for fuzzy or approximate matching, which is required to account for differences in formats and data structures. 
In our case, using encryption techniques (like secure multi-party computation~\cite{yao1982protocols} or homomorphic encryption~\cite{hesurveyimplementation}) for privacy preservation disallow the use of direct or equality comparisons as there is no support for such operators over \emph{encrypted} records. Operators, in general, including arithmetic, bit, logical, and others are limited in these tools. 

\noindent\textbf{Contributions}
This paper presents \textbf{AMPPERE}, an abstract computation model for performing privacy-preserving entity resolution offering platform-agnostic, algorithmic abstractions using variants of Jaccard similarity~\cite{jaccard1901etude}. The provided abstractions are \emph{universal}: they support widely used arithmetic and vector operations and first-class support for private matrix manipulation. The abstraction model is presented as a pipeline that can be instantiated via simple privacy-preserving operations. We demonstrate the feasibility of \textbf{AMMPERE}'s abstraction model by transparently implementing it atop the secure multi-party computation platform Sharemind~\cite{sharemind} and the open-source lattice-based homomorphic encryption library PALISADE~\cite{PALISADE}. We undertake a rigorous evaluation of our system which demonstrates the efficacy of our abstraction model, and a precise characterization of the induced overheads resulting from the \emph{strong cryptographic protections} provided by \textbf{AMPPERE}. 

\noindent\textbf{Roadmap} The paper is structured as follows: \cref{sec:problem-definition} algorithmically defines the ER \& privacy-preserving ER problems as well as evaluation metrics;
%and \cref{sec:challenges} calls out the challenges for privacy preservation, thus motivating the need for an abstract platform-agnostic machine for universal privacy-preserving ER;
the rest of \cref{sec:preliminaries} introduces the technical machinery that form the algorithmic components of \textbf{AMPPERE}; \cref{sec:approach} describes and analyzes \textbf{AMPPERE}, the operators supported, and how it can be adapted to concrete implementations. \Cref{sec:experiments} reports the settings, results, and findings of our experiments; \cref{sec:related-works} lists  prior approaches to PPER, and finally \cref{sec:conclusion-future-work} presents our conclusions and directions for further work.

%% file: preliminary.tex
\newcommand\tab[1][1cm]{\hspace*{#1}}
\section{Algorithmic definition and tools}\label{sec:preliminaries}
\input{problem.tex}

We now introduce our main algorithmic tools for entity resolution: the set similarity metric of Jaccard similarity~\cite{jaccard1901etude} and its derivative MinHashLSH~\cite{leskovec2014}, which is a MinHash-based variant of Local Sensitive Hashing (LSH) designed for searching similar documents from a large corpus. We also introduce the two main cryptographic and privacy tools that we leverage in our implementations of \textbf{AMPPERE}: secure multi-party computation (MPC)~\cite{yao1982protocols} and homomorphic encryption (HE) ~\cite{hesurveyimplementation}.

%\label{sec:preliminaries}
\subsection{Jaccard similarity}
Jaccard similarity~\cite{jaccard1901etude} is a well-known and widely used set-based similarity metric, commonly used for feature generation and similarity measurement in data mining and information retrieval tasks.
Given two sets of tokens $x$ and $y$, Jaccard similarity refers to the size of the intersection divided by the size of the union of the token-sets, that is, $Sim_{Jaccard}(x, y) = \frac{|x \cap y|}{|x \cup y|} = \frac{|x \cap y|}{|x| + |y| - |x \cap y|}$.

Some existing PPER approaches~\cite{sehili2015privacy} extend the Jaccard similarity metric to  privacy-preserving settings by encrypting input sets into bit arrays (or fingerprints), and comparing them using Tanimoto similarity~\cite{tanimoto1958elementary}, which computes similarity in a manner similar to Jaccard: $Sim_{Tanimoto}(x, y) = \frac{|x \wedge y|}{|x \vee y|} = \frac{|x \wedge y|}{|x| + |y| - |x \wedge y|}$, where $|x|$ and $|y|$ denotes the number of set bits (cardinality) in the bit array $x$ and $y$ respectively. 
% In data mining and information retrieval, Jaccard similarity is a commonly used technique for feature generation and similarity measurement. 
We use Jaccard similarity to perform ER by selecting a threshold $t$ such that $(r_1, r_2) \in M$, if $Sim_{Jaccard}(r_1, r_2) > t$.

\subsection{MinHashLSH}
 MinHash exploits \textit{min-wise independent} permutations to generate similarity-preserving signatures for each record. The probability of getting two similar MinHash signatures from two permutations of sets is identical to applying Jaccard directly over the them; formally, $Pr[h_{min}(perm(x)), \allowbreak h_{min}(perm(y))]\allowbreak =\allowbreak Sim_{Jaccard}(x,y)$ where $h_{min}$ denotes MinHash and $perm$ denotes the random permutation function. The LSH algorithm then splits the signature into smaller bands (chunks) of a fixed length where only the records that share at least one band are considered to be potentially similar (i.e. they have partially common signatures). MinHashLSH Blocking ~\cite{leskovec2014} is an inverted-index inspired method utilizing MinHashLSH bands as blocking keys and associated record ids as values.
 
 MinHashLSH is statistically robust and efficient for finding similar records or documents; since blocking keys are signatures, they do not reveal information relating to original records. For these reasons, we use it as our blocking method of choice in \cref{sec:blocking}.

\subsection{Privacy tools}
\noindent\textbf{Secure multi-party computation} (MPC)~\cite{yao1982protocols} ensures that a set of mutually distrusting parties involved in the same distributed computation correctly derive the result of the computation, even when a subset of parties are dishonest, without revealing any new information other than the result of computation. Formally, if the parties are $p_1, p_2, \cdots, p_n$ and their private data is $d_1, d_2, \cdots, d_n$, a MPC protocol defines the public function $F$ which computes the result over the private data, i.e., $F(d_1, d_2, \cdots, d_3)$ while preserving the secrecy of $d_1, d_2, \cdots, d_n$.

% Sharemind~\cite{sharemind} is a client-server MPC platform that supports black-box general-purpose MPC computation for programs written in a domain-specific language, SecreC. Sharemind supports several different MPC schemes called \emph{protection domains}. Our implementation (\Cref{sec:sharemind}) leverages the \texttt{shared3p} protection domain, which stands for 3-out-of-3 secret sharing with passive security, and uses an \emph{additive} secret sharing scheme. This typically results in the following client-server state machine: a SecreC program comparing inputs of two data sources $A$ and $B$, locally transforming input $A$ (and resp. $B$) into a $3$-way \emph{share} where each share is sent to the Sharemind application server. Note that none of the shares alone, nor any pair of these shares give any information about the original secret input value. The $3$ Sharemind servers perform a MPC protocol, and as long as all $3$ are not colluding, the output of the computation is returned to the client without any of the servers learning anything about the private input of the data sources.

\noindent\textbf{Homomorphic Encryption}
\label{sec:homomorphic-encryption}
(HE) is a public-key cryptographic scheme that enables certain computations (e.g. addition, multiplication) to be performed directly over encrypted data, without a need for decryption~\cite{hesurveyimplementation}. Conceptually, it can be defined as an algorithmic quadruple of ($KeyGen$, $E$, $D$, $f$): $KeyGen$ produces a public, private key pair $(p_k, s_k)$; $E$ is an encryption algorithm that uses $p_k$ to encrypt a given plaintext $p_i$, outputting a corresponding ciphertext $c_i$; $f$ is an evaluation function to be directly evaluated over ciphertexts $c_1 \cdots c_n$, outputting $c_f$ representing the encrypted computation; $D$ is a deterministic algorithm that takes in $c_f$, decrypts it using $s_k$, and outputs the result: $f(p_1 \cdots p_n)$. The most important component of HE is it's \emph{homomorphic property}: suppose an $f$ that operates on inputs $p_k$ \&  $c_1$ $...$ $c_n$, and outputs $c_f$. For all $p_1 \cdots p_n$, it must hold that if $p_f$ = $p_1 \circ p_2$, $c_1 = E(p_k, p_1)$, and $c_2 = E(p_k, p_2)$, then $Prob(D(f(s_k, c_1, c_2)) \neq{p_f}$ is \textit{negligible} (where $\circ$ refers to a homomorphic operation) ~\cite{hetheoryapplication}. 

\noindent \textbf{Threshold HE} 
Threshold homomorphic encryption combines the use of threshold cryptography ~\cite{Desmedt2011} and HE to provide functionality via MPC. Here, $n$ different parties interact, initially, to compute a joint public and private key, where the private key is split into $n$ secret shares, one per party. Each party encrypts their own data using the public key, and the encrypted result is then decrypted, collectively, by some subset of parties assuming that a majority of them are not malicious~\cite{10.1007/3-540-44987-6_18}. 

In our two implementations of \textbf{AMPPERE}, we use Sharemind~\cite{sharemind-sdk}, a client-server platform for black-box general-purpose MPC computation, and PALISADE~\cite{PALISADE}, an open-source lattice-based HE (\& threshold-HE) library.
% For parameter settings and experiments, refer to \cref{sec:palisade-impl} and \cref{sec:experiments}.

%% file: problem.tex
\subsection{Problem definition}
\label{sec:problem-definition}
% 0.5 page.
We formalize the \textbf{ER} problem as a triple $P=(D_1,\allowbreak D_2,\allowbreak M)$ where $D_1$ and $D_2$ are two different datasets consisting of a collection of unique records. $M$ denotes the matching record pairs between the two datasets, that is, $M=\{(r_i, r_j)\,|\,r_i = r_j, r_i \in D_1, r_j \in D_2\}$ where $r_i=r_j$ indicates that $r_i$ and $r_j$ refer to the same entity in the real world. 
% \pe{This definition is misleading: it should say that $r_i$ and $r_j$ refer to the same entity in the real world.} 
With \textbf{PPER}, we still deal with $D_1$, $D_2$, and $M$, however, we additionally say the probability that the data owners of $D_1$, $D_2$ learn any information outside of what record pairs exist in $M$ is negligible, and the probability that an adversary $A$ learns anything, including $M$ or any part of $M$, is negligible. 

Additionally, we define a \textbf{full comparison} between $D_1$ and $D_2$ to be $T=\{(r_i, r_j)\,|\,r_i\in D_1, r_j \in D_2\}$, and the number of candidate pairs to be $|T|$, the Cartesian product between sizes of $D_1$ and $D_2$ (i.e. $|D_1| \times |D_2|$). 
% Naturally, the non-matching record pairs can be denoted as $N$. 
We focus on token-based entity resolution algorithms for finding $M$, where each record is represented by a set of unique tokens $r = (t_1, t_2, \cdots)$. 
% The quality and performance of these linking algorithms is measured by \textbf{precision} ($pr$) and \textbf{recall} ($re$). Denoting the pair-set found by the algorithm to be $P$, $pr$ = $\tfrac{M \cap P}{P}$ and $re$ = $\tfrac{M \cap P}{M}$. For comprehensive estimation of performance, \textbf{F-score} = $\frac{2 \cdot pr \cdot re}{pr \cdot re}$ is used to harmonically combine these two scores.
\textbf{Blocking} algorithms prune $T$ by removing pairs that are unlikely to be part of the solution $M$. 
\textit{Inverted-index} blocking algorithms are popular methods that use a function $\beta$ to encode each record into  blocking keys $\beta(r)=(k_1,k_2,\cdots)$, preserving \textit{candidate record pairs} that share at least one $k$. Therefore, the pruned candidate pair set is defined as $T'=\{(r_i, r_j)\,|\,|\beta(r_i) \cap \beta(r_j)|>0, r_i \in D_1, r_j \in D_2\}$.
A good blocking algorithm reduces as many candidate pairs as possible without losing truly matched pairs~\cite{michelson2006}. We use \textbf{pairs completeness} ($PC$) to measure the percentage of true pairs that are blocked, and \textbf{reduction ratio} ($RR$) to measure how well the method reduces the number of candidate pairs. These are formally defined as: $PC=\tfrac{M \cap T'}{M}$, $RR = 1 - \tfrac{|T'|}{|T|}$, and $F\textnormal{-}score=\tfrac{2 * PC * RR}{PC + RR}$.

% 0. plaintext data can not be directly distributed to computing host.
% 1. can not simply use one-way hash algorithms, records from different sources may have different formats or representations (also vulnerable to statistical attack)
% 2. obfuscating data or doing fuzzy data structure based matching loses accuracy / linking quality.
% 3. with MPC or FHE, the comparison can not be directly over encrypted records which means you can not directly test if two records or two tokens are the same. operators, including arithmetic, bit and logical, are limited.
% 4. many MPC protocols are specifically designed for certain problems, no general solution (though it's because of the efficiency purpose)
% 5. scalability.

%% file: approach.tex
%
%\vspace{-5.3mm}
\begin{figure}[!h]
    \includegraphics[width=0.98\linewidth]{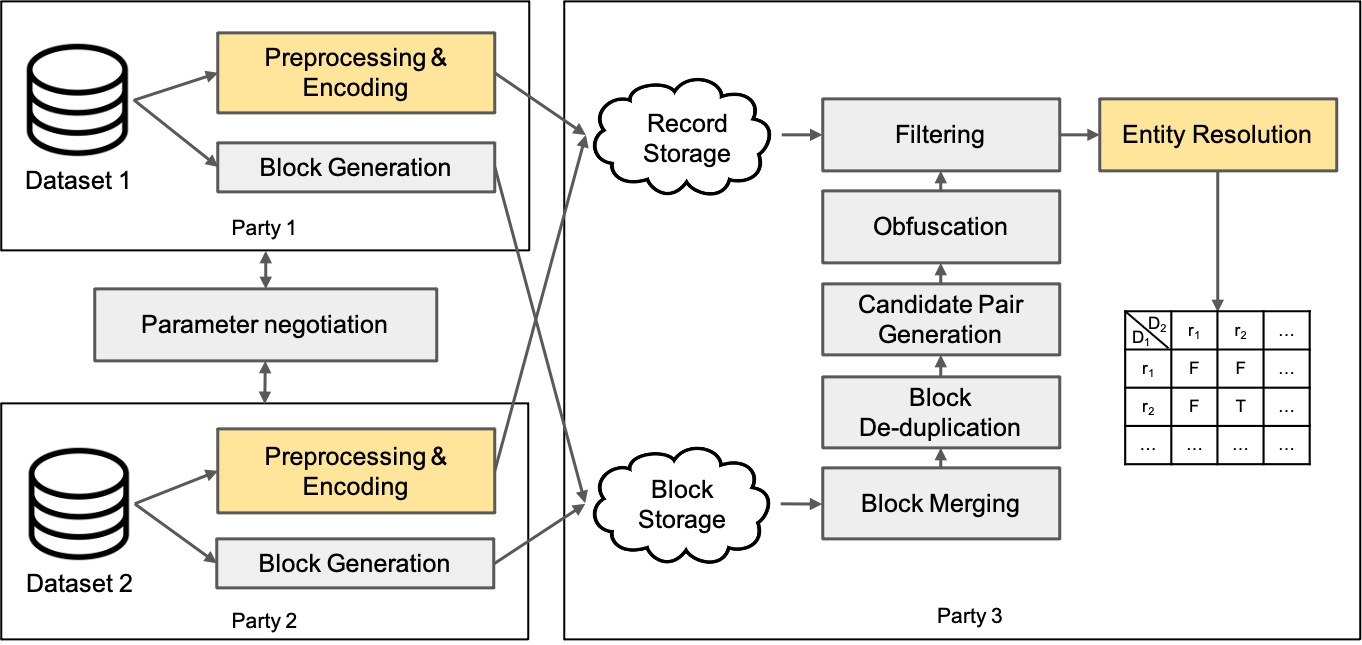}
    \caption{Abstraction model}
    \label{fig:abstraction-model}
    % \vspace{-0.15in}
\end{figure}
\section{AMPPERE}
\label{sec:approach}
We propose \textbf{AMPPERE}, a platform-agnostic abstract machine for privacy-preserving entity resolution. We first introduce the specific challenges regarding PPER, present our computation model, and then two concrete implementations.
% : one based on a secure multi-party computation platform, and the other on a homomorphic encryption library.

\noindent\textbf{\underline{Challenges of PPER}}
\label{sec:challenges}
Privacy preservation adds another dimension to the entity resolution problem, making it much more challenging and computationally difficult. Keeping in mind our original example, hospital 1 \& 2, contrary to regular ER approaches, cannot directly compare their records to determine that records for Peter and Bruce overlap between their databases. In the privacy-preserving setting, their records need to firstly be encrypted, and distributed to a trusted computation host who can perform the data matching. Once encrypted, however, there is limited support in computing equality, similarity or joins, in terms of general-purpose functionality and scalability. Furthermore, one-way hashing techniques are rendered ineffective as they are susceptible to statistical-based attacks, and different data sources may have different representations making it hard to apply them.

% The rest of this section presents how \amppere addresses these challenges while still providing strong cryptographic privacy guarantees for ER.

\subsection{Abstraction model}
Our model is presented such that it can be instantiated using only simple platform-agnostic, privacy-preserving primitive operations. As shown in \cref{fig:abstraction-model}, both dataset providers (i.e. $P_1$, $P_2$) negotiate on parameters first (e.g., Jaccard, blocking threshold), then apply encoding on each of their records, individually, while also computing blocks. Encoded records and blocks are then uploaded and stored in a cloud owned by a third party (i.e $P_3$), where blocks are merged and deduplicated, and candidate pairs are filtered. ER (privately) is performed on the remaining qualified pairs. The accuracy, in performing PPER, of our abstractions and algorithms is 100\% given correct implementations, as evidenced by \cref{sec:sharemind} \& \cref{sec:palisade-impl}.

\begin{figure*}[!t]
    \includegraphics[width=0.85\linewidth]{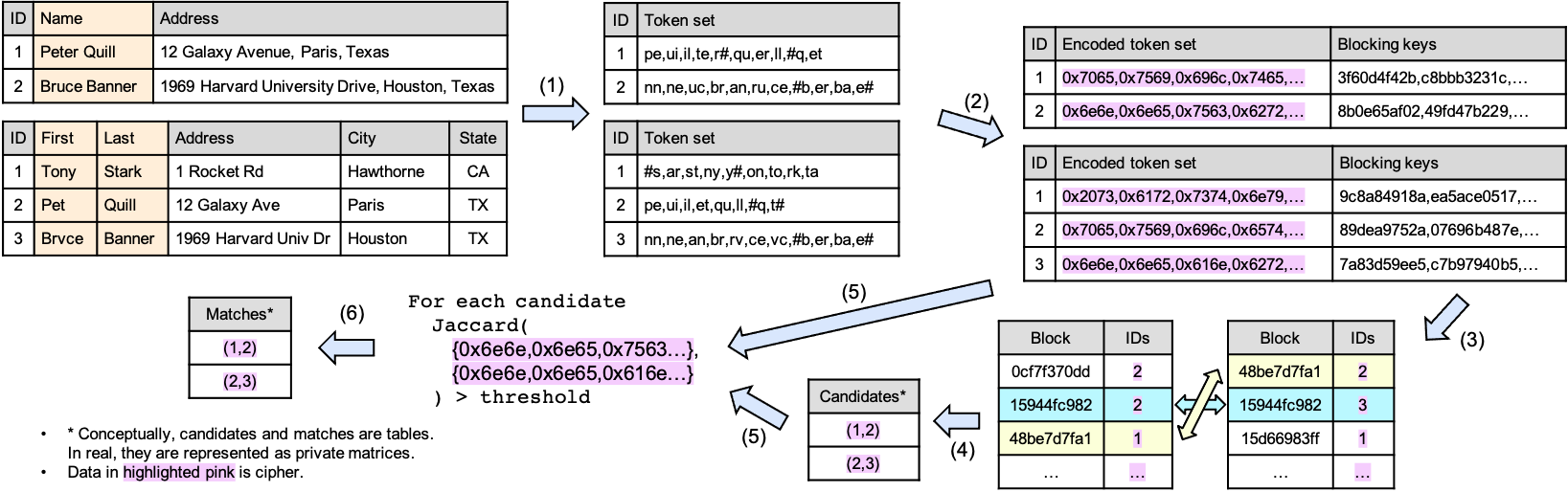}
    \caption{Example of data flow. Original records with selected attributes after (1) pre-processing and tokenization to construct token sets. The token sets are then (2) ASCII-encoded, blocking keys are generated and both are encrypted and (3) sent to computational unit. Blocks with same keys are merged and (4) de-duplicated candidate pairs are generated. (5) Jaccard similarity is applied to the candidate pairs and (6) matches are found.}
    \label{fig:example}
    % \vspace{-0.15in}
\end{figure*}

\begin{table}[ht]
    % \vspace{-0.1in}
    \centering
    \caption{Abstraction Model Operators. (E) indicates the primitive has an element-wise version.}
    % \vspace{-0.15in}
    \begin{tabular}{| p{0.3 \linewidth} | p{0.6 \linewidth} | }
    \hline
    \textbf{Category}  & \textbf{Name} \\ \hline 
    Primitive operators & $\mathtt{(E)\{Add,Sub,Mul\}, DotProduct}$, $\mathtt{\{L,R\}Shift, Size, Transpose, Enc/Dec}$ \\ \hline
    Set intersection size & Pairwise join, Vector rotation, Vector extension, Sorting, Matrix join \\ \hline
    Ternary operations & $\mathtt{Choose\{,Vec,VecExt,Mat,MatExt}\}$ \\ \hline 
    Private matrix manipulation & $\mathtt{MaskGen, Vector\{Lookup,Update\}}$, $\mathtt{Matrix\{Lookup,Update\}}$ \\ \hline 
    \end{tabular}
    \label{tab:operators}
    % \vspace{-0.1in}
\end{table}

\subsubsection{Core operations}
\label{sec:core_operations}

Our abstraction model relies on several core operations, listed in \cref{tab:operators}, that should be defined first. The model assumes that the following \textbf{primitive operators} for vector and matrix data structures are available: \texttt{Add} for addition, \texttt{Sub} for subtraction, \texttt{Mul} for multiplication, and their corresponding element-wise versions: \texttt{EAdd/ESub/EMul}. Additionally, \texttt{DotProduct} for computing dot products, \texttt{LShift/RShift} for shifting vectors left and right, \texttt{Size} for getting the size of a vector or matrix, and \texttt{Transpose} for transposing a vector or matrix. \texttt{Enc} and \texttt{Dec} are encryption \& decryption operators. Additional parameters (e.g. cryptographic keys, etc.) may be required, but are omitted here for simplicity. Optionally, logic and bit operators may be used wherever applicable, but we provide workarounds using only arithmetic operations.

The first essential step in computing Jaccard similarity is calculating \textbf{set intersection size}: the intersection size of two tokenized-record sets. We list and compare several possible approaches using the primitive operators mentioned above; choosing the appropriate approach depends on the implementation of the operations that it relies on. We use $v_1$, $v_2$ to indicate two input token-sets (vectors) that contain only unique entries.
\begin{itemize}
    \item \textbf{Pairwise join} (PJ) Compare entries in $v_1$ and $v_2$ in a pair-wise manner while accumulating the results into a counter. Since this method must be computed serially, it is inefficient and does not allow the use of optimized primitive operators.
    \item \textbf{Vector rotation} (VR) Fix the longer of the two input vectors (e.g. $v_2$), pad the smaller (e.g. $v_1$) with arbitrary, non-integer elements not a part of $v_1$ or $v_2$ until the sizes are equal (if necessary), and count the number of overlapping elements represented by zeroes in $v_1-v_2$. This calculation must be done $Size(v_1)$ steps as each iteration requires a right shift of $v_2$ by one entry. For example, say $v_1=[1, 2]$ and $v_2=[2, 2, 3]$, then $v_1'=[1, 2, *]$ and $v_2$ remains unchanged. Computing $v_1'-v_2$ in the first iteration returns $[-1, 0, -3]$, giving us one 0, repeat this process for $Size(v_1)$ iterations. The total number of 0's represents the number of overlapping elements. Note, this method is an adaptation of basic PSI (private set intersection) protocol presented by Chen et al ~\cite{chen_psi_acm}.
    
    \item \textbf{Vector extension} (VE) Extend $v_1$ to $v_1'$ by repeating the it $Size(v_2)$ times, and extend $v_2$ to $v_2'$ by repeating each element of $v_2$ $Size(v_1)$ times; the total number of 0's in $v_1'-v_2'$ is the number of overlapping elements. For example, say $v_1=[1,2,4]$ and $v_2=[2,3]$, then $v_1'=[1,2,4,1,2,4]$ and $v_2'=[2,2,2,3,3,3]$, hence, $v_1'-v_2'=[-1,0,2,-2,-1,1]$ and the overlap is 1. This approach offers an alternative to VR (in terms of space vs. time complexity) by pre-allocating memory for all possible entry-wise comparisons and only needing one direct comparison (subtraction) between the two resulting vectors. 

    \item \textbf{Sorting} (SO) Merge $v_1$ and $v_2$ into $v_3$, sort $v_3$, and then count overlaps between two derived vectors from $v_3$: the first not including the first entry, and the second not including the last. For example, say $v_1=[1,2,4]$, $v_2=[2,3]$, and $Sorted(v_3)=[1,2,2,3,4]$; then computing $[1,2,2,3]-[2,2,3,4]=[-1,0,-1,-1]$ gives us an overlap of 1.
    
    \item \textbf{Matrix join} (MJ) Compute the join between $v_1^{T}$ and $v_2^{T}$ and count the size of the result.
\end{itemize}

Control flow statements are involved in almost all programs. Choice (e.g. if) and loop (e.g. for, while) based flows require conditional expressions for execution; in privacy-preserving settings, however, encrypted Boolean values cannot be used in conditional operations in order to avoid exposing the execution path, unless they are decrypted. \textbf{Ternary operations} can be designed to bypass this limitation: they take a condition followed by two values; if the condition is true, the first value is returned, otherwise the latter (e.g. \texttt{if/else} or  \texttt{:?}). If the two values have the same data type and dimension, the return value can be determined by pure arithmetic. We design a set of oblivious-style ternary operators~\cite{ohrimenko2016oblivious,constable2018liboblivious} which work with encrypted conditions and inputs in various forms (numbers, vectors and matrices) in \cref{alg:tenary-operator}.

\begin{algorithm}[t]
\small
\caption{Ternary Operators}
\label{alg:tenary-operator}
    % Set Function Names
    \SetKwFunction{FEMul}{EMul}
    \SetKwFunction{FEAdd}{EAdd}
    \SetKwFunction{FESub}{ESub}
    \SetKwFunction{FSize}{Size}
    
    \SetKwFunction{FChoose}{Choose}
    \SetKwFunction{FChooseVec}{ChooseVec}
    \SetKwFunction{FChooseVecExt}{ChooseVecExt}
    \SetKwFunction{FChooseMat}{ChooseMat}
    \SetKwFunction{FChooseMatExt}{ChooseMatExt}
    
    \SetKwInput{KwInput}{Input}
    \KwInput{$cond$: bool, $n_1, n_2$: numbers, $v_1, v_2$: vectors, $m_1, m_2$: matrices}
    
    % Write Function with word ``Function''
    \SetKwProg{Fn}{Function}{:}{}
    \Fn{\FChoose{$cond$, $n_1$, $n_2$}}{
        \KwRet $cond * (n_1 - n_2) + n_2$\;
    }
    \BlankLine
    
    \SetKwProg{Fn}{Function}{:}{}
    \Fn{\FChooseVec{$cond$, $v_1$, $v_2$}}{
        \KwRet $\FEAdd(\FEMul(cond, \FESub(v_1, v_2)), v_2)$\;
    }
    \BlankLine
    
    \SetKwProg{Fn}{Function}{:}{}
    \Fn{\FChooseVecExt{$cond$, $v_1$, $v_2$}}{
        $cond: vector_{\FSize(v_1)} = cond$\;
        \KwRet $\FChooseVec(cond, v_1, v_2)$\;
    }
    \BlankLine
    
    \SetKwProg{Fn}{Function}{:}{}
    \Fn{\FChooseMat{$cond$, $m_1$, $m_2$}}{
        \KwRet $\FEAdd(\FEMul(cond, \FESub(m_1, v_2)), m_2)$\;
    }
    \BlankLine
    
    \SetKwProg{Fn}{Function}{:}{}
    \Fn{\FChooseMatExt{$cond$, $m_1$, $m_2$}}{
        $cond: matrix_{\FSize(m_1)} = cond$\;
        \KwRet $\FChooseMat(cond, m_1, m_2)$\;
    }
\end{algorithm}

We also define functionality for getting \& setting values for encrypted vectors and matrices. We refer to the set of these operations as \textbf{private matrix manipulation}, described in \cref{alg:matrix-manipulation}. To privately access vectors and matrices, arithmetic operations need to run over a special mask generated from an encrypted index ($idx$, $rol$, $col$). In \texttt{MaskGen} (Line 1-3), \texttt{EEq} element-wisely compares if $seq$ and $idx$ are equal, and sets the corresponding element in the result to be 1 if equal and 0 if not. This procedure generally uses a logical operator, but can also be achieved via arithmetic operations as follows: $(seq-idx)*\frac{-1}{seq-idx+\xi}+1$ where $\xi$ is a small offset added to prevent dividing by zero ($seq$, $idx$ must be integers). If a division operator is not available, an element $n$ can be multiplied by a randomly-sampled encrypted plaintext element $r$, generated by $P_1$ or $P_2$, i.e. $enc(r)*enc(nx)=enc(r*n)$. $P_3$ can decrypt this product and compute its reciprocal, $\frac{1}{r*n}$, in clear. Finally, it can encrypt this value and multiply with $enc(r)$ to output the reciprocal of $n$: $enc(\frac{1}{r*n})*enc(r)=enc(\frac{1}{n})$. 
% \yi{$ -(c_1-c_2) / max(1, c_1-c_2) + 1$ where $max(c_1,c_2)=(c_1+c_2+|c_1-c_2|)/2$, $c_1$ and $c_2$ are 0 or positive integers}
% \yi{Fermat little theorem also works here https://crypto.stackexchange.com/questions/58047/equality-check-in-homomorphic-encryption, but the scheme needs to support modulo operator} \ta {also can use floor operation for only integer arithmetic: scaling factor to be $f$, the formula is $\lfloor-f(c1-c2) / (f(c1-c2)-1)\rfloor + 1$, where f=$10^n$ and $n>=1$ } \yi{it could also be achieved by pure bit-wise operations, e.g., xor and bit shift, need to think about this}

\begin{algorithm}[t]
\small
\caption{Private matrix manipulation}
\label{alg:matrix-manipulation}
    % Set Function Names
    \SetKwFunction{FMul}{Mul}
    \SetKwFunction{FSize}{Size}
    \SetKwFunction{FEnc}{Enc}
    \SetKwFunction{FInnerProduct}{InnerProduct}
    \SetKwFunction{FTranspose}{Transpose}
    \SetKwFunction{FChooseVec}{ChooseVec}
    \SetKwFunction{FVectorLookup}{VectorLookup}
    \SetKwFunction{FChooseMat}{ChooseMat}
    
    \SetKwFunction{FVectorLookup}{VectorLookup}
    \SetKwFunction{FVectorUpdate}{VectorUpdate}
    \SetKwFunction{FMatrixLookup}{MatrixLookup}
    \SetKwFunction{FMatrixUpdate}{MatrixUpdate}
    
    \SetKwFunction{FMaskGen}{MaskGen}
    \SetKwFunction{FEEq}{EEq}
    
    \SetKwInput{KwInput}{Input}
    \KwInput{$size$, $idx$, $row$, $col$: integers, $vec$: vector, $mat$: matrix, $val$: number}

    \SetKwProg{Fn}{Function}{:}{}
    \Fn{\FMaskGen{$size$, $idx$}}{
        $seq: [0,1,\cdots,size]$, $idx: vector_{size} = idx$\;
        \KwRet $\FEEq(seq, idx)$\;
    }
    \BlankLine
    
    \SetKwProg{Fn}{Function}{:}{}
    \Fn{\FVectorLookup{$vec$, $idx$}}{
        $mask \gets \FMaskGen(\FSize(vec), idx)$\;
        \KwRet $\FInnerProduct(vec, \FTranspose(mask))$\;
    }
    \BlankLine
    
    \SetKwProg{Fn}{Function}{:}{}
    \Fn{\FVectorUpdate{$vec$, $idx$, $val$}}{
        $val: vector_{\FSize(vec)} = val$\;
        $mask \gets \FMaskGen(\FSize(vec), idx)$\;
        \KwRet $\FChooseVec(mask, val, vec)$\;
    }
    \BlankLine
    
    \SetKwProg{Fn}{Function}{:}{}
    \Fn{\FMatrixLookup{$mat$, $row, col$}}{
        $mask \gets \FMaskGen(\FSize(mat)[0], row)$\;
        $row\_vec \gets \FMul(mask, mat)$\;
        \KwRet $\FVectorLookup(row\_vec, col)$\;
    }
    \BlankLine
    
    \SetKwProg{Fn}{Function}{:}{}
    \Fn{\FMatrixUpdate{$mat$, $row, col$, $val$}}{
        $val: matrix_{\FSize(mat)}=val$\;
        $row\_mask \gets \FMaskGen(\FSize(mat)[0], row)$\;
        $col\_mask \gets \FMaskGen(\FSize(mat)[1], col)$\;
        $mask \gets \FMul(\FTranspose(row\_mask),col\_mask)$\;
        \KwRet $\FChooseMat(mask, val, mat)$\;
    }
\end{algorithm}

% Here, the index $idx$/$row$/$col$ is in plaintext, but to not reveal the position of an encrypted element in a vector or matrix to an executor, the index of the elements also need to be obfuscated \pe{we are using several words as encrypted, de-mystified, obfuscated, if they all mean encrypted, we should use only one term} on the dataset provider's end. A simple method is to construct a mapping that maps the ciphertext of each element to an unique random identifier in the range of the size of the vector or matrix. We call this mapping function \texttt{Hash} and denote the name as \texttt{Hash1} and \texttt{Hash2} for two datasets respectively.
% For some libraries may end up with different cipher text even the plain text is the same, a two-step mapping can be constructed by first map the cipher text of each element to its original value then further map it to the unique random number. Notice that the "intermediate step" in the mapping is no need to be delivered to the third party.

\subsubsection{Entity resolution}
\label{sec:entity-linking}
We model two datasets, $D_1$ and $D_2$ (e.g. $D_i=[Enc([t_1, t_2, \cdots])), \cdots]$), each containing encrypted records (e.g. $Enc(r) \in D_1$) where each record is formed by a set of tokens (e.g. $t \in r_1$). To represent records as token-sets, we use $n$-gram~\cite{banerjee2003design} tokenization, a general-purpose, fault-tolerant method to tokenize each record with ASCII encoding, represented in 8 bits. We choose a $n=2$ (bi-gram) encoding, therefore each token occupies 16 bits allowing us to use any data type that has more than 16 bits as a container. In particular, we use the 64 bit container (e.g. \texttt{uint64}), so 4 encoded tokens to be placed in one, as 4 falls $\in [2,8]$, the commonly-used range of n-grams~\cite{lesher1999effects}.

The input for entity resolution, besides $D_1$ and $D_2$, is a threshold $t$, and the output is a $|D_1| \times |D_2|$ sized encrypted, Boolean matrix $results$ whose row-indices indicate record ids from $D_1$, and column-indices indicate record ids from $D_2$; a cell's value indicates  
if the corresponding records are in the solution $M$. We use Jaccard similarity to compute how \textit{similar} two records are based on their tokens, and only consider a match if the similarity is greater than $t$. 
We thereby define the similarity for a candidate record pair $(r_1, r_2)$ to be $\frac{inter}{Size(r_1) + Size(r_2) - inter} + \epsilon > t$ where $inter$ indicates the set intersection size of two token-sets and $\epsilon$ is a value added for tolerating the accuracy of floating point representation.

\subsubsection{Blocking}
\label{sec:blocking}

Blocks are represented using a hashmap where each key is a blocking key, and each value is an encrypted vector containing record ids; that is, $\{"bkey1": Enc([rid_1, rid_2, \cdots]), \cdots\}$. Blocks are generated by each party ($P_1, P_2$) individually, and are subsequently merged by $P_3$ if they have the same blocking key. 
% The record ids in each block are paired to be "candidate pairs" if they indicate the records they represent come from different datasets. 
Record ids in a block coming from different datasets are grouped into candidate pairs.

\Cref{alg:er-with-blocking} shows how blocks are used in our entity resolution pipeline. Lines 1-10 represent block merging and deduplication. Block deduplication refers to the process of determining which encrypted record id pairs have the same blocking key, but are from different datasets; this is done to avoid duplicated candidate pairs and duplicate comparisons. Our approach uses an encrypted Boolean matrix $cand\_pairs$ with the same dimensions \& indices as $results$, where True values denote candidate pairs.
This provides a simple solution to the difficult problem of constructing privacy-preserving inverted indices for deduplication ~\cite{wang2015inverted}. \texttt{VectorLookup} retrieves record ids in encrypted form, and \texttt{MatrixUpdate} privately labels candidate pairs.
In order to execute entity resolution on the candidate pairs, the program needs to get the record ids of each pair in clear.
Even if there exists a method that is able to fetch records and invoke filtering with encrypted record ids, it is inevitable that an attacker could sniff memory or time the code block to detect if certain record pairs are faked. Our solution to not leak the total number of candidate pairs and certain combinations of record pairs is to obfuscate the candidate pair matrix before decryption. In Line 11, the $cand\_pairs$ matrix is converted to $obfu\_pairs$ by adding (union-ing) random truthy values with \texttt{AddNoise} before it is decrypted. The $results$ matrix is then created along with the $dummies$ matrix (Line 12-13). After applying filtering and ER (Line 14-17), the results are privately \emph{chosen} from $results$ or $dummies$ via \texttt{ChooseMatExt} with the help of $cand\_pairs$ (Line 18), and the added noise is automatically eliminated. 
If a random number generator for \texttt{AddNoise} is not supported on $P_3$, a random matrix can be pre-generated and encrypted by $P_1$ or $P_2$ for obfuscation. 

% Because decrypting the values in the encrypted $cand\_pairs$ matrix is not possible, looping to compute similarity on record pairs directly is not an option. As an alternative, the algorithm first adds a dummy record at the end of $D_1$ and $D_2$ respectively (Line 11). After this, it calls \texttt{MatrixLookup} with encrypted indices, and based on the returning encrypted value, either a real record id pair or the dummy id pair will be returned (Line 15). Finally, using this id pair, we can fetch back the record token sets and Jaccard similarity can be computed (Line 16 - 17). This works as \texttt{Jaccard} will return instantly if it sees two empty input sets. \cref{fig:example} illustrates this process from the perspective of data flow for the hospital records example.

\begin{algorithm}[t]
\small
\caption{PPER with blocking}
\label{alg:er-with-blocking}
    \SetKw{Continue}{continue}
    \SetKwFunction{FSize}{Size}
    \SetKwFunction{FEnc}{Enc}
    \SetKwFunction{FDec}{Dec}
    \SetKwFunction{FChooseVecExt}{ChooseVecExt}
    \SetKwFunction{FVectorLookup}{VectorLookup}
    \SetKwFunction{FMatrixUpdate}{MatrixUpdate}
    \SetKwFunction{FMatrixLookup}{MatrixLookup}
    \SetKwFunction{FJaccard}{Jaccard}
    \SetKwFunction{FAddNoise}{AddNoise}
    
    \SetKwInput{KwInput}{Input}
    \SetKwInput{KwOutput}{Output}
    \KwInput{$D_1$ and $D_2$ are two datasets, $b_1$ and $b_2$ are two blocks, $t$ is the Jaccard similarity threshold.}
    \KwOutput{$results$ is a matrix.}
    
    \BlankLine
    
    \tcp{Merging and de-duplication}
    $cand\_pairs: matrix_{(\FSize(D_1),\FSize(D_2))} = \FEnc(false)$ \;
    \For{$bkey \in b_1.keys$}{
        \If{$bkey \not\in b_2.keys$}{
            \Continue;
        }
        
        $r_1\_ids \gets b_1[bkey]$, $r_2\_ids \gets b_2[bkey]$\;
        
        \For{$r_1\_id \gets 0$ \KwTo $\FSize(r_1\_ids)$}{
            $enc\_r_1\_id = \FVectorLookup(r_1\_ids, r_1\_id)$\;
            \For{$r_2\_id \gets 0$ \KwTo $\FSize(r_2\_ids)$}{
                $enc\_r_2\_id = \FVectorLookup(r_2\_ids, r_2\_id)$\;
                $cand\_pairs = \FMatrixUpdate(cand\_pairs,$ \
                    $enc\_r_1\_id, enc\_r_2\_id, \FEnc(true)$\;
            }
        }
    }
    
    \BlankLine
    
    \tcp{Obfuscation}
    $obfu\_pairs \gets \FDec(\FAddNoise(cand\_pairs))$ \;
    
    \BlankLine
    
    \tcp{Filtering and entity resolution}\
    $results: matrix_{(\FSize(D_1),\FSize(D_2))} = \FEnc(false)$ \;
    $dummies: matrix_{(\FSize(D_1),\FSize(D_2))} = \FEnc(false)$ \;
    
    \For{$i \gets 0$ \KwTo $\FSize(D_1)$}{
        \For{$j \gets 0$ \KwTo $\FSize(D_2)$}{
            \If{$obfu\_pairs[i,j]$}{
                $results[i,j] \gets \FJaccard(D_1[i], D_2[j], t)$\;
            }
        }
    }
    
    $results \gets \FChooseMatExt(cand\_pairs, results, dummies)$ \;
\end{algorithm}

\subsubsection{Privacy analysis}
This section analyzes the privacy properties of \amppere from end to end. 
Firstly, we note that \amppere provides cryptographic protections for privacy-preservation: the MPC-based implementation provides privacy of computation to all parties subject to a \emph{majority} being corrupted~\cite{KatzLindell2014} whereas the HE-based implementation provides privacy against a \emph{honest-but-curious} adversary. Next, we note that $P_1$, $P_2$ first encode their input records and perform blocking on their end, and then send their corresponding encrypted data (records \& record ids) to $P_3$. Blocking keys are sent to $P_3$ in clear. 
%The blocking is executed by party 1 and party 2, then blocking keys in clear and encrypted record id vectors are submitted to party 3. 
The output from $P_3$ is an encrypted matrix where $P_1$, $P_2$ are responsible for finally decrypting it to retrieve the \emph{matching} record pairs. 
$P_3$ needs the vector sizes to calculate the size of the set intersection, resulting in disclosure of the encoded record sizes.
%In the computing process of party 3, since calculating the set intersection size requires acquiring the size of vector, encoded record size is disclosed. 
Block merging, deduplication, candidate pair generation and filtering use ternary \& private matrix operations, all which need to invoke the \texttt{Size} primitive.
Consequently, the total number of blocks, the size of each block, and the size of each dataset are disclosed. 
%For block merging, de-duplication, candidate pair generation and filtering, which rely on ternary operators and private matrix manipulations need to invoke primitive \texttt{Size}, so the total number of blocks, the size of each block and the size of each dataset are revealed. 
Therefore, after a complete run, $P_1$, $P_2$ learn the size of each dataset and the matched record id pairs only.
% \pe{what does it mean to learn the mis-matched record id pairs?}
$P_3$ learns the size of each dataset, the size of each encoded record, the size of each block and all blocking keys. No record content is disclosed and because of that no useful information is revealed to $P_3$. Additionally, the blocking key is the partial MinHash signature, which is irreversible; the total number of blocks as well as the size of each block also do not provide any valuable information. Strictly, two blocks with the same key will be merged, but $P_3$ only knows the size of each block before merging. If two typical blocks with the same key $k$ are $b_1^k$ and $b_2^k$, the maximum possible matches could be found in merged block is $min(|b_1^k|, |b_2^k|)$. 
Records often appear in multiple blocks, making it difficult for $P_3$ to estimate the number of matches. 
The total number of candidate pairs is also protected. Because of obfuscation, its value is somewhere between $|T'|$ to $|T|$, making it such that an attacker is not able to identify which pairs are noisy and which are real.
%Hence, it is almost not possible for party 3 to track or precisely estimate the number of matches.
%\pe{"Almost not possible" is meaningless
% One thing to notice is that, the decryption of result matrix can also be done by party 3. If so, party 3 learns all matched and mis-matched record id pairs but party 1 and 2 only learn the matched record id pairs.

\subsection{Sharemind implementation}
\label{sec:sharemind}
SecreC~\cite{secrec} is a C-like domain-specific programming language designed for the Sharemind platform application \cite{randmets2017programming}. SecreC provides generic programming facilities, hiding details of its underlying secure protocol; it has many differences as compared to general-purpose languages:
%\begin{itemize}
    (i) SecreC distinguishes between public and private data types at the system level via a security type (separate from the data type). The programmer must indicate whether certain data is public or private, and then the implementation can automatically convert one-way from public to private if necessary. 
    (ii) Operations on private data types are limited. For example, \textit{bool}, \textit{int}, \textit{uint} and \textit{float} do not support bitwise \texttt{and} and bitwise \texttt{or} operators.
    (iii) Private values cannot be used in conditional statements.
    (iv) No file system access. 
    The host machine and running environment can exchange data only via input arguments, via keydb, a key value store based on Redis, or via a vector-based table\_database; the data has to be batched and pre-uploaded through command line arguments to these containers.
    % \pe{Why is this a restriction? Is there something that you can do with files that you cannot do with these mechanisms? If not it is not a restriction, it may be an inconvenience, but that is different}
%\end{itemize}

% SecreC provides \texttt{upload} and \texttt{link} programs and  C++ interfaces to \pe{say what these are for}
%Two SecreC programs and corresponding C++ interface are built in Sharemind implementation: \texttt{upload} and \texttt{link}. 
Our implementation has two components: \texttt{upload} and \texttt{link}. \texttt{upload} loads encoded records and blocks to the Sharemind keydb instance and \texttt{link} is used for entity resolution.
%Because Sharemind does not support direct file system access, encoded records and blocks are uploaded to the keydb instance through the \texttt{upload} program. 
The key for a record is a public string in the form \texttt{\{dataset id\}\textunderscore \{record id\}}, and the value is a private \texttt{uint64} array which stores encoded tokens. The format of a block is similar to a record, but the key has an additional prefix as a namespace. For a record, if the associated blocking key exists, the program will fetch the corresponding vector back and merge the record id into it with the \texttt{cat} function; 

The blocking keys belonging to $D_1$ are acquired from the keydb instance, and are constructed with the blocking keys with $D_2$'s identifier; the blocks belonging to different datasets, but with the same blocking key are fetched, candidate pairs are generated and deduplicated, and finally stored into $cand\_pairs$ with \texttt{matrixUpdate}. 
After this, the \texttt{link} program obfuscates the candidate pairs, and runs filtering and ER. Utilizing Sharemind's API, we implemented the following methods for computing set intersection size: Pairwise join, Vector extension (with internal function \texttt{gather\textunderscore uint64\textunderscore vec}), Sorting (with \texttt{quicksort}) and Matrix join (with \texttt{tableJoinAes128}). 
In the end, $results$ is reshaped to a 1D array (as Sharemind does not allow matrices to be returned) and is shipped back to the host.
% Notice that Sharemind runtime environment needs three nodes, the compiled SecreC program has to be deployed onto all these servers.

\subsection{PALISADE implementation}
\label{sec:palisade-impl}

%As mentioned in \cref{sec:homomorphic-encryption}, PALISADE provides implementations for various HE schemes, lattice cryptography and related cryptographic primitives in C++. 
Our PALISADE implementations work with the Brakersi-Gentry-Vaikuntanathan (BGV) \cite{cryptoeprint:2011:277} (for integer arithmetic) and the Cheon-Kim-Kim-Song (CKKS) \cite{heforarithmeticofapproximatenumbers} (for approximate numbers) constructions, in the form of their respective RNS (residue number system) variants \cite{heevaloftheaescircuit}\cite{rnsvariantforapproximatehe}. As a first step in utilizing any scheme within PALISADE, one must instantiate a corresponding \texttt{CryptoContext} object for encryption functionality. Each context must be configured with relevant parameters that provide certain bit-security levels and performance; in our implementations we configured the following: \emph{multiplicative depth}, \emph{scale factor bits}, \emph{plaintext modulus}, \emph{batch size}, \emph{key switching technique}, and \emph{security level}. Both BGV and CKKS are \emph{fully} homomorphic (FHE) in the sense that they both support (an arbitrary, unbounded number of) additive and multiplicative homomorphic operations~\cite{fhegentrythesis}. In PALISADE, these schemes work as parametrized, \textit{leveled} versions, meaning computations are computed up to a pre-determined depth. To establish a fair comparison between both schemes, we keep their corresponding ring dimensions and CRT moduli ~\cite{cryptoeprint:2019:939} the same. \Cref{tab:palisade-parameter} displays our value choices for the parameters above; further fine-tuning can be done for potentially more efficient runtimes. 
\begin{table}[ht]
    % \vspace{-0.1in}
    \caption{BGV \& CKKS Parameters. (B) indicates parameter for blocking.}
    % \vspace{-0.15in}
    \label{tab:palisade-parameter} % is used to refer this table in the text
    \begin{tabular}{| p{0.23\linewidth} | p{0.12\linewidth} | p{0.20\linewidth} | }
        \hline 
        \textbf{Parameter} & \textbf{BGV} & \textbf{CKKS} \\ [0ex] % inserts table
        %heading
        \hline %inserts double horizontal lines
        ring dim. & 4096 & 4096, 8192 (B) \\ % inserting body of the table 
        \hline
        mult. depth & 1 & 1, 2 (B) \\ \hline
        scale factor bits & - & 40  \\ \hline
        plaintext mod. & 65537 & - \\ \hline
        batch size & - & 16 \\ \hline
        key switching & BV & BV \\ \hline
        security level & 128 bits & 128 bits \\ \hline % [1ex] adds vertical
    \end{tabular}
    \vspace{0.1in}
\end{table}

We developed 4 specific implementations in PALISADE: 
a two party protocol in both BGV and CKKS respectively, and their corresponding threshold (MPC) variants. In the two party protocol, $P_2$ encrypts their data, sends it to $P_1$ who then operates with $P_2$'s encrypted data, and their own data (in clear). In the MPC version, we apply threshold HE where $P_1$ and $P_2$ must first agree on a public/private key-pair to be used for encryption, and distributed decryption. Then, both parties encrypt and send their data to $P_3$ to carry out the computation over both of their (encrypted) data. Parties are modeled via individual data-structures, each referencing public/private keys generated via the \texttt{KeyGen} function and are realized as separate entities through different threads that take on the evaluation compute. We implemented the Pairwise join, Vector extension, and Vector rotation set intersection size approaches within PALISADE. PJ encodes each token of each record as a separate \texttt{Ciphertext} object with the value of the integer located at slot 0, whereas VR and VE are able to take advantage of \textit{plaintext packing} (i.e. how many plaintexts you can fit into a single ciphertext slot) by encoding an entire record of tokens together as one ciphertext (Single Instruction Multiple Data (SIMD) computations). PJ computes the ER process in serial, and only relies on the PALISADE function \texttt{EvalSub} to homomorphically compare each token of one record, pairwise with each token of another record. VR and VE both utilize \texttt{EvalAtIndex} to homomorphically rotate (right shift) along with \texttt{EvalSub} for subtractions, while VE additionally uses \texttt{EvalSum} to repeat elements for a certain \textit{batchSize} over the ring dimension. 

We implemented blocking in the CKKS scheme. A discussion around why we chose CKKS vs. BGV for blocking is located in \cref{sec:palisade-performance}. We used our own custom HE implementation of ternary operators (\cref{alg:tenary-operator}) and private matrix manipulation (\cref{alg:matrix-manipulation}) for deduplication and filtering (\cref{alg:er-with-blocking}).
% Due to the additional computational overhead here, the ring dimension is increased to \numprint{8192}, as opposed to \numprint{4096} for the non-blocking approaches. 

%% file: experiment.tex
\section{Experiments}
\label{sec:experiments}
% 2.25 page.

\subsection{Datasets and settings}
%In this subsection, we demonstrate the structure and feature of the dataset that is used in experiments and how it can be generated. We also show the expected performance of running Jaccard similarity and MinHashLSH blocking on this dataset. Since the experiments are on two different implementations, the configurations and settings of servers and programs are provided.

This section describes the dataset used for evaluation and the configurations used for our two implementations.

\noindent \textbf{Febrl dataset}
Febrl is a synthetic dataset generated using \textit{dsgen}~\cite{christen2005probabilistic}, a tool to produce census records including a customizable amount of noise. We generated it using the following settings: Maximal number of duplicates for one original record: 5, Maximum number of modifications per field: 5, Maximum number of modifications per record: 5, Probability distribution for duplicates: zipf, Type of modification: typo, ocr, phonetic.
% We prepared 2 different sizes of this dataset in 100, 1k and each of them is split into a 20\% part for $D_1$ and 80\% part for $D_2$.
Our dataset used for experiments contains 100 records, split into a 20\% component for $D_1$ and 80\% component for $D_2$. Since our purpose of experimentation is to demonstrate the feasibility, not scalability of \textbf{AMPPERE}, we did not generate any larger datasets.
% \pe{This sentence doesn't make sense}\yi{Should we just remove this sentence or any suggestions? If we don't mention this, readers may ask why the test datasets are only 20x80.}
We further processed our dataset using a Python script to create bi-gram, integer tokens for each record, and to compute MinHashLSH blocking keys for each record with thresholds ranging from 0.1 to 0.9 with step 0.1.
%The original dataset is then being pre-processed by a Python script which bi-gram tokenized and encoded each record into integer tokens. 
%In addition, the script also computed corresponding MinHashLSH blocking keys for each record with threshold ranging from 0.1 to 0.9 with step 0.1.

\begin{figure}[!t]
  \includegraphics[width=\linewidth]{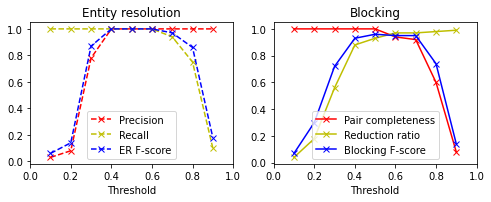}
  \caption{Expected entity resolution and blocking results}
  \label{fig:eval-er-blocking}
%   \vspace{-0.15in}
\end{figure}

\noindent\textbf{Expected performance}
To better understand the dataset, we created a Python script to apply the same blocking and ER methods in a non-privacy preserving setting, computing precision, recall, $PC$, $RR$ and $F$-$score$. 
\Cref{fig:eval-er-blocking} shows that ER precision and recall reach 1.0 when the threshold is in range 0.4 to 0.6. $PC$ remains 1.0 when the threshold is $\leq$ to 0.5 and $RR$ climbs shapely in that range. 

\begin{figure}[!t]
  \includegraphics[width=0.6\linewidth]{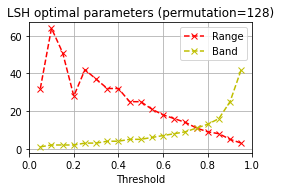}
  \caption{Optimal MinHashLSH blocking key size}
  \label{fig:eval-minhash-optimal}
%   \vspace{-0.15in}
\end{figure}

We used Datasketch~\cite{datasketch} to calculate MinHash signatures for records and to generate LSH blocking keys. We set the number of permutations to 128, and the relative importance of false positive/negative probability to 0.5 each. Datasketch calculates \texttt{range} and \texttt{band} to minimize the weighted sum of probabilities of false positives/negatives with all range and band combinations under a given threshold. The optimal blocking key sizes (the same as \texttt{range}) in different thresholds are shown in \cref{fig:eval-minhash-optimal}. The blocking key size is correlated with the $b$: the blocking threshold.

% \begin{figure}[ht]
%     \begin{minipage}{0.63\linewidth}
%       \includegraphics[width=\linewidth]{resources/eval_er_blocking.png}
%       \caption{Expected entity resolution and blocking results}
%       \label{fig:eval-er-blocking}
%     \end{minipage}\hfill
%     \begin{minipage}{0.33\linewidth}
%       \includegraphics[width=\linewidth]{resources/eval_minhash_optimal.png}
%       \caption{Optimal MinHashLSH blocking key size}
%       \label{fig:eval-minhash-optimal}
%     \end{minipage}
% \end{figure}

\noindent\textbf{Environment settings}
The experiments on our Sharemind and PALISADE implementations are conducted on three virtual Ubuntu 18.04.4 LTS servers each with 2 CPUs from Intel Xeon CPU E5-2690 v4 @ 2.60GHz and 4GB memory. All servers are in the same network and the average PING latency is around 0.12-0.23ms.

The Sharemind runtime environment~\cite{sharemind-sdk} and implementations are deployed on all three servers, as for the SecreC program the \texttt{shared3p} private domain requires distributed across 3 different nodes. For the fairest comparison of evaluation results with Sharemind, we used the following configurations in PALISADE: compiled the PALISADE cmake project with the \texttt{-DWITH-NATIVEOPT} flag set to on; used single-threaded runtimes to establish baselines between systems, configured via \texttt{export OMP-NUM-THREADS=1}; turned CPU scaling off. We also disabled obfuscation (for both platforms) to achieve the most optimal running time possible.

\subsection{Entity resolution performance}
% 0.5 page + several images.
% Compare the performance of different set intersection approaches on Sharemind and Palisades.
We tested the efficiency of executing Jaccard similarity with the different approaches for computing set intersection size described in \cref{sec:core_operations}. For each approach, we recorded the execution time of Jaccard similarity on 100 randomly sampled single pairs.

\subsubsection{Sharemind performance}
\begin{figure*}[!t]
    \centering
    \begin{subfigure}[b]{0.33\textwidth}
        \includegraphics[width=\textwidth]{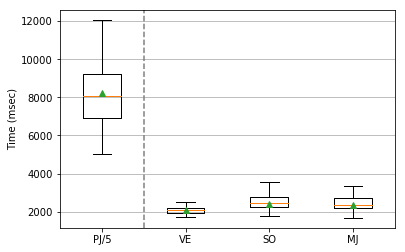}
        \caption{Sharemind}
        \label{fig:eval-sharemind-jaccard}
        % \vspace{-0.15in}
    \end{subfigure}
    \hfill
    \begin{subfigure}[b]{0.33\textwidth}
        \centering
        \includegraphics[width=\textwidth]{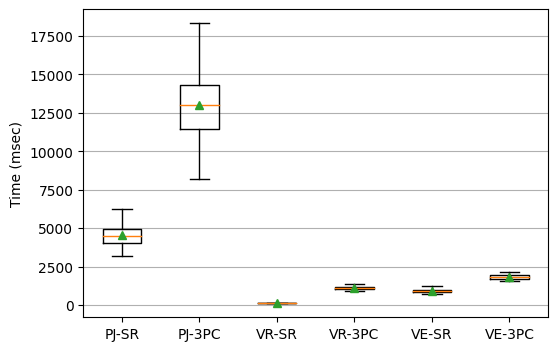}
        \caption{PALISADE: BGV}
        \label{fig:eval-eval_palisade_jaccard_bgvrns}
        % \vspace{-0.15in}
    \end{subfigure}
    \hfill
    \begin{subfigure}[b]{0.33\textwidth}
        \centering
        \includegraphics[width=\textwidth]{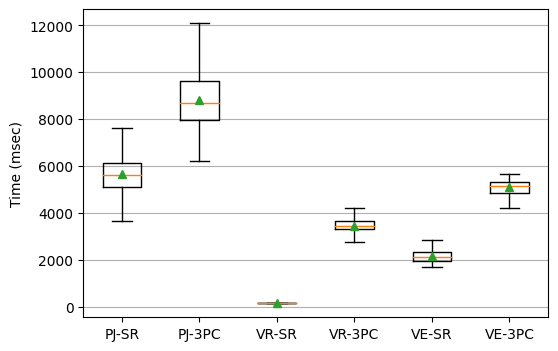}
        \caption{PALISADE: CKKS}
        \label{fig:eval-eval_palisade_jaccard_ckks}
        % \vspace{-0.15in}
    \end{subfigure}
    \caption{Time cost (per candidate pair) for PPER}
    \label{fig:eval-er}
    % \vspace{-0.15in}
\end{figure*}

In Sharemind, we adopted PJ, VE, SO and MJ respectively. Since PJ is extremely slow compared with other approaches, the time cost of all its runs have been divided by 5 in order to be shown properly in one diagram. \Cref{fig:eval-sharemind-jaccard} shows that except for PJ, which takes on average 8k ms per record pair, the other three approaches have around the same performance: around 1.8k-4k ms per record pair. PJ executes the two loops serially and does not benefit from any parallelism of vector operations. VE achieved the best efficiency, likely a result of Sharemind's \texttt{gather\textunderscore uint64\textunderscore vec} internal method which saves DSL level operations. The shortcoming of VE, however, is that if the input records were long, overhead due to memory allocation could be a problem. The results of SO show more variance because quick sort is not a stable sorting algorithm; MJ is almost identical to SO. Notice that MJ uses Sharemind's \texttt{tableJoinAes128} function which requires the input records to be encoded in \texttt{xor\_uint32} and this limitation halves the capacity of input containers (which is \texttt{uint64}). Overall, if the average length of records is not long, VE and MJ can be good choices, otherwise SO is better suited.

\subsubsection{PALISADE performance}
\label{sec:palisade-performance}
In PALISADE, we adopted the PJ, VR, and VE method in both BGV and CKKS. The runtime efficiency here is better than Sharemind's because these experiments are not distributed across three machines. We perform the experiments in two modes: sender-receiver or $SR$ (2 party protocol) and three-party computation or $3PC$ (threshold variant). The VR and VE approaches both performed significantly faster than PJ, as reflected in \cref{fig:eval-eval_palisade_jaccard_bgvrns} and \cref{fig:eval-eval_palisade_jaccard_ckks}, with VR being the fastest of the three. VR is able to reach speeds of ~125-175 ms per record pair for both BGV and CKKS in the sender-receiver variation and between 1.5 and 4 seconds in the the corresponding threshold (3PC) version. VR and VE are much faster than PJ because they use \textit{plaintext packing} as described in \cref{sec:palisade-impl}. Additionally, VR is faster than VE because VR only relies on primitive operations (rotations and subtractions), while VE requires extending records using an expensive operation for repeating elements in encrypted form (\texttt{EvalSum}). 
% So even though, theoretically, VE appears to be a faster approach (it tries to optimize time complexity while increasing space complexity), with our implementation in PALISADE, VR is able to perform PPER more efficiently.
Consistent with Sharemind, the PJ approach computes ER serially and does not benefit from any parallelization or packing. There are other ways to implement VE and VR in PALISADE that might perform faster or slower, depending on what you try to optimize (time or space). 

\noindent \textbf{BGV vs. CKKS Performance} BGV generally had equal or better runtime efficiency as compared to CKKS, which is expected. The main difference between the two schemes is that CKKS is an approximation algorithm that supports fast rescaling while dealing with fixed-precision arithmetic. BGV, depending on implementation, either does not support this type of rescaling or requires bootstrapping, which is a much more expensive operation. Since BGV works with exact numbers, when numbers are multiplied, they should not wrap around the \textit{plaintext modulus}, whereas in CKKS, rescaling is supported after every multiplication. Since our non-blocking approaches did not require a \textit{multiplicative depth} greater than 1 (because there was no need for sequential homomorphic multiplications), BGV becomes an ideal choice; however, when nested-multiplications are involved, BGV becomes quickly inefficient as its ciphertext modulus for depth $k$, must be at least $p^k$, where $p$ is also a power of $2^k$. CKKS is the ideal choice in these situations as it can rescale each time, keeping the parameters small.

\noindent \textbf{Takeaway} Determining the optimal approach for privacy-preserving entity resolution may vary with the computational platform and the concrete implementation, but our experiments above show many options are applicable and reasonably efficient.

\begin{figure*}[!ht]
    \centering
    \begin{subfigure}[bc]{0.33\textwidth}
        \includegraphics[width=\textwidth]{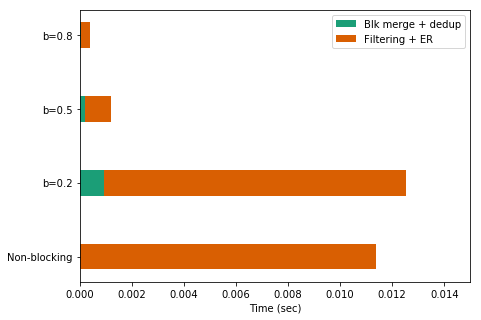}
        \caption{Non-privacy preserving}
        \label{fig:eval-normal-blocking}
        % \vspace{-0.15in}
    \end{subfigure}
    \hfill
    \begin{subfigure}[bc]{0.33\textwidth}
        \includegraphics[width=\textwidth]{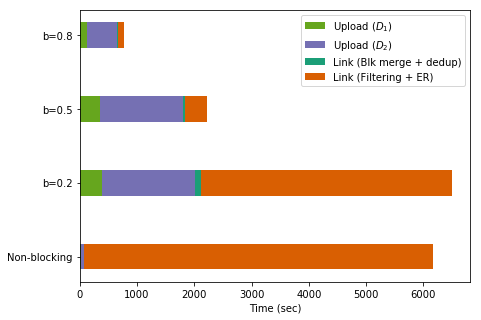}
        \caption{Sharemind}
        \label{fig:eval-sharemind-blocking}
        % \vspace{-0.15in}
    \end{subfigure}
    \hfill
    \begin{subfigure}[bc]{0.33\textwidth}
        \includegraphics[width=\textwidth]{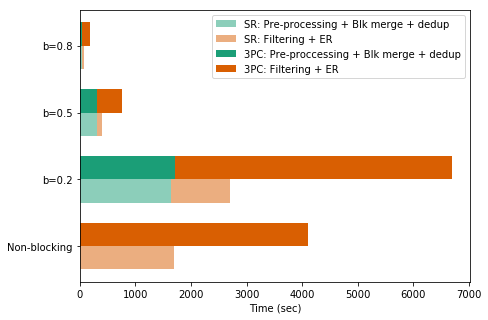}
        \caption{PALISADE: CKKS}
        \label{fig:eval-palisade-blocking}
        % \vspace{-0.15in}
    \end{subfigure}
    \caption{Overall time cost for PPER with blocking}
    \label{fig:eval-blocking}
    % \vspace{-0.15in}
\end{figure*}

\subsection{Blocking performance}
% 1 page.
% Time cost of the system with or without blocking. 1-2 images.
In this experiment, we show the effectiveness of blocking in performing (private) entity resolution with the two implementations. We used the non-blocking approach as the baseline (performs all \numprint{1600} pairwise comparisons) and ran blocking with a MinHashLSH threshold $t$ of t=0.2, 0.5, and 0.8. The blocking key sizes corresponding to these threshold values are 28, 25 and 9 (resp.) and the numbers of distinct comparisons after blocking are limited to \numprint{1314}, 118 and 36 (resp.) as compared to the overall \numprint{1600}.

\Cref{fig:eval-sharemind-blocking} demonstrates the total and step time cost of PPER with blocking in Sharemind. The baseline takes around \numprint{6000} seconds to make all \numprint{1600} pair-wise similarity comparisons, whereas, for the blocking version, the dataset upload cost depends on the size of blocking keys. Hence, when $t=0.8$, it takes the least amount of time to upload data as there are only 9 keys per record and block merging, deduplication and candidate pair generation costs are minor. In the entity resolution step, when $t=0.2$, it needs \numprint{1314} comparisons, which while smaller than the baseline, shows an overall longer runtime due to extra overhead incurred via data upload and block operations. When $t=0.5$, the number of comparisons along with execution time drops drastically, and this trend continues when $t=0.8$, but with recall loss as a trade-off cost.

As with Sharemind, blocking in PALISADE shows great time cost savings as it filters out unnecessary pairwise record comparisons. \Cref{fig:eval-palisade-blocking} shows the overall time cost resulting in PALISADE with the various blocking thresholds against the non-blocking baseline. We implement blocking in both the $SR$ and $3PC$ variants with CKKS as opposed to BGV due to the optimized performance in dealing with rescaling after multiplications, and because of good support for scalar/plaintext \& ciphertext operations in implementing ternary operators presented in \cref{alg:tenary-operator}.
Unlike Sharemind, we perform blocking in PALISADE we three main steps: preprocessing (generating block data structures), block merging + deduplication + candidate pair generation, and filtering + ER calculations. In \cref{fig:eval-palisade-blocking}, we group the runtime of steps 1 and 2 into one bar as, for all blocking threshold levels, preprocessing remains $<1$ second. For both $SR$ and $3PC$ variants, when $t=0.2$, the overall time-cost is actually longer than the baseline (just like with Sharemind), because while filtering is applied through deduplication and candidate pair generation, it only removes around 300 overall comparisons, which does not outweigh the overhead incurred by those steps. When $t=0.5, 0.8$, blocking and filtering is able to brunt much of the computation, by limiting the number of comparisons needed drastically, allowing for the ER calculation to be as minimal as possible.

\noindent \textbf{Takeaway} The incurred cost of privacy preservation is significant and evident as shown in \cref{fig:eval-normal-blocking}. Filtering and ER takes between 0 and ~0.013 seconds as compared to our PPER approaches: between ~\numprint{800} and ~\numprint{7000} seconds in Sharemind,
between ~100 and ~\numprint{3000} seconds in the $SR$ variant, and between ~200 and ~\numprint{7000} seconds in the $3PC$ variant in PALISADE.

%% file: relatedwork.tex
\vspace{-1mm}
\section{Related work}
\label{sec:related-works}
ER has been a well-studied area for many decades, from specific matching algorithms including Jaccard to semi/full automatic frameworks. PPER, however, has only been investigated recently. Vatsalan et al. \cite{vatsalan2013taxonomy,vatsalan2017privacy} provide a taxonomy of some privacy-preserving record linkage techniques, and suggest a general architecture. Some earlier work has explored~\cite{dusserre1995one,quantin1998ensure} exploiting secure one-way hashing and masking algorithms to protect the content of records, but this is limited as it does not allow for \emph{similar} variants of records. Some algorithms that rely on certain probabilistic data structures (e.g. Bloom filters~\cite{schnell2009privacy, niedermeyer2014cryptanalysis}), though, do exist, and provide reasonable performance \& privacy protection, while also maintaining relative similarity between records. Additionally, some record linking approaches depend on the modification of string matching algorithms (e.g. private edit distance~\cite{wang2015efficient,zhu2017efficient}, P4Join~\cite{sehili2015privacy}). With the increase of word/sentence embedding techniques, some methods including \cite{scannapieco2007privacy, bonomi2012frequent} convert original records into multi-dimensional embedding vectors which still preserve some principle characteristics, and can be used in similarity comparison and clustering.

Very limited exploration~\cite{lindell2005secure}, however, of MPC or HE based approaches exists today, and what does exist only handles the case of \textit{exact matching}~\cite{emekci2006privacy}, due to expensive computation and lack of computational tool support. Currently, there is no general and comprehensive computational model like \amppere for privacy-preserving blocking and ER. In recent times, though, many excellent MPC frameworks and HE libraries including Microsoft SEAL~\cite{sealcrypto}, HELib~\cite{helib}, Obliv-C~\cite{zahur2015obliv} have been developed and are becoming more applicable for large-scale computing purposes. It is clear and evident that a general, precise, and \emph{scalable} solution for ER based on these techniques will shine and break through in the near future.

%% file: conclusion.tex
% \vspace{-3mm}
\section{Conclusion and future work}
\label{sec:conclusion-future-work}
% 0.25 page. Future work is about 1) finding the optimal threshold for Jaccard and MinHashLSH (Ground truth sampling, PP-annotation) 2) Schema awareness (schema alignment, no need to be PP). 3) FHE CKKS tuning and distributed communication over a network for palisades

We proposed an abstract machine for PPER and adopted it to two concrete implementations (Sharemind \& PALISADE). Our experiments showed that both implementations could produce results with the exact or relatively equivalent correctness to traditional ER methodologies, while \emph{maintaining} privacy. This lends credence to the possibilities of solving ER problems with MPC and HE. 

Under this direction, many possible and promising future works can be explored. For instance, how to tune and find the optimal parameters for the matching algorithm without revealing data via a privacy-preserving adaptation of ground truth sampling and/or oracle annotation mechanisms. Moreover, data-set schema and metadata can also be utilized to further improve the accuracy and performance of matching results. And finally, scaling efficiently to larger and more complex data-sets and real-world scenarios is also a non-trivial problem that can be explored, but this will become increasingly computationally feasible as MPC and HE implementations become more efficient.